\documentclass[twocolumn,english,final,showpacs,pra,lengthcheck]{revtex4}
\usepackage{ae,aecompl}
\usepackage[T1]{fontenc}
\usepackage[latin9]{inputenc}
\usepackage{color}
\usepackage{babel}
\usepackage{amsmath}
\usepackage{amssymb}
\usepackage{graphicx}
\usepackage[unicode=true,pdfusetitle,
 bookmarks=true,bookmarksnumbered=false,bookmarksopen=false,
 breaklinks=false,pdfborder={0 0 1},backref=false,colorlinks=true]
 {hyperref}

\makeatletter
\@ifundefined{textcolor}{}
{%
 \definecolor{BLACK}{gray}{0}
 \definecolor{WHITE}{gray}{1}
 \definecolor{RED}{rgb}{1,0,0}
 \definecolor{GREEN}{rgb}{0,1,0}
 \definecolor{BLUE}{rgb}{0,0,1}
 \definecolor{CYAN}{cmyk}{1,0,0,0}
 \definecolor{MAGENTA}{cmyk}{0,1,0,0}
 \definecolor{YELLOW}{cmyk}{0,0,1,0}
}

\usepackage{textcomp}
\usepackage{babel}

\usepackage{amsfonts}

\providecommand{\U}[1]{\protect\rule{.1in}{.1in}}
\makeatletter
\providecommand{\LyX}{L\kern-.1667em\lower.25em\hbox{Y}\kern-.125emX\@}
\@ifundefined{textcolor}{}
{
\definecolor{BLACK}{gray}{0}
\definecolor{WHITE}{gray}{1}
\definecolor{RED}{rgb}{1,0,0}
\definecolor{GREEN}{rgb}{0,1,0}
\definecolor{BLUE}{rgb}{0,0,1}
\definecolor{CYAN}{cmyk}{1,0,0,0}
\definecolor{MAGENTA}{cmyk}{0,1,0,0}
\definecolor{YELLOW}{cmyk}{0,0,1,0}
}
\makeatother
\makeatother

\makeatother

\usepackage{babel}

\makeatother

\begin{document}

\preprint{This line only printed with preprint option}

\title{Concatenated dynamical decoupling with virtual pulses}

\author{Gonzalo A. \'Alvarez}

\altaffiliation{gonzalo.alvarez@tu-dortmund.de}

\affiliation{Fakult\"at Physik, Technische Universit\"at Dortmund, Dortmund,
Germany.}

\author{Alexandre M. Souza}

\altaffiliation{alexandre@e3.physik.uni-dortmund.de}

\affiliation{Fakult\"at Physik, Technische Universit\"at Dortmund, Dortmund,
Germany.}

\author{Dieter Suter}

\altaffiliation{Dieter.Suter@tu-dortmund.de}

\affiliation{Fakult\"at Physik, Technische Universit\"at Dortmund, Dortmund,
Germany.}


\begin{abstract}
The loss of quantum information due to interaction with external degrees
of freedom, which is known as decoherence, remains one of the main
obstacles for large-scale implementations of quantum computing. Accordingly,
different measures are being explored for reducing its effect. One
of them is dynamical decoupling (DD) which offers a practical solution
because it only requires the application of control pulses to the
system qubits. Starting from basic DD sequences, more sophisticated
schemes were developed that eliminate higher-order terms of the system-environment
interaction and are also more robust against experimental imperfections.
A particularly successful scheme, called concatenated DD (CDD), gives
a recipe for generating higher order sequences by inserting lower
order sequences into the delays of a generating sequence. Here, we
show how this scheme can be improved further by converting some of
the pulses to virtual (and thus ideal) pulses. The resulting scheme,
called vCDD, has lower power deposition and is more robust against
pulse imperfections than the original CDD scheme.
\end{abstract}
\maketitle

\section{\textit{\emph{Introduction}}}

Quantum information processing has acquired a huge interest over the
last decades. It can potentially solve many problems qualitatively
faster than classical information processing. The quest to implement
this scheme has led to a lot of progress on quantum control methodologies
and technologies (see, e.g., \cite{Ladd2010}). The main obstacle
for its implementation is the sensitivity of quantum systems to interactions
with external degrees of fredom that destroy or modify the information
to be processed in an uncontrolled way \cite{Zurek2003}. A number
of techniques are currently being developed to make reliable quantum
computing possible in the presence of environmental noise. A relatively
simple technique is dynamical decoupling (DD) \cite{Viola1999,Viola2003,Khodjasteh2005,Uhrig2007,Gordon2008,Uhrig2009,West2010,Khodjasteh2010,Clausen2010,Yang2010,Wang2011a,Quiroz2011,Kuo2011},
which uses sequences of control pulses applied to the system qubits.
This technique does not require any overhead in terms of ancilla qubits
and requires no additional controls over those that are already needed
for information processing. This field has seen significant progress
over the last years, and the concept has been demonstrated on a number
of different systems \cite{Cywinski2008,Yang2008,Biercuk2009a,Du2009,Clausen2010,Alvarez2010,Lange2010a,Barthel2010,pasini_optimized_2010,Ryan2010,Ajoy2011,Souza2011,Almog2011,Bardhan2011,Pan2011,Shukla2011,Bluhm2011,Naydenov2011}. 

In the limit of infinitely many ideal refocusing pulses, the DD scheme
allows one to completely eliminate the decoherence due to the environmental
noise. However, in any real physical implementation, the control pulses
necessarily have finite duration and unavoidable imperfections. This
leads to a significant reduction of the DD performance, and the effect
of a real pulse sequence on the system can actually reduce the fidelity
instead of improving it \cite{Khodjasteh2007,Alvarez2010,Hodgson2010,Souza2011a,Wang2011,Xiao2011,Wang2010,Khodjasteh2011,Peng2011}.
These recent results have shown that efficient DD schemes must be
able to preserve the system fidelity even in the presence of non-ideal
control fields \cite{Alvarez2010,Ryan2010,Souza2011,Khodjasteh2011,Souza2011a,Souza2011b,Cai2011}.

One strategy that was shown to be robust against imperfections is
a technique called concatenated dynamical decoupling (CDD), which
is based on a building block sequence that is concatenated recursively
\cite{Khodjasteh2005,Khodjasteh2007}. This procedure improves the
DD performance with the concatenation order. If the delays between
the pulses can be reduced indefinitely, CDD was demonstrated to improve
its performance with the concatenation order. However if the delays
between the pulses are constrained or the pulses have errors, it was
predicted \cite{Khodjasteh2007} and experimentally demonstrated \cite{Alvarez2010}
that an optimal concatenation order exists, and beyond that the DD
performance will not improve or even deteriorate.

To increase the concatenation order, the procedure inserts the lower-order
CDD sequence within the delays of the building block sequence. If
the pulses have imperfections and the buiding block sequence compensates
partially their effects at the end of the cycle, the CDD sequence
will also compensate them at the end of the complete sequence. However,
if the average delay between pulses is kept fixed \cite{Alvarez2010,Ajoy2011,Souza2011},
the duration of the CDD cycle increases exponentially with the CDD
order. The compensation of the pulse imperfections only occurs at
the end of the cycle. If the cycle time exceeds the correlation time
of the environmental fluctuations, this error compensation becomes
inefficient and the DD performance decreases.

In this article, we present a new approach to the CDD scheme that
does not require waiting for the end of the cycle to compensate the
pulse imperfections. Instead, they are always compensated over the
duration of the lowest order cycle. This is done by introducing virtual
pulses for the building block sequence to generate the higher CDD
order. Being virtual, \emph{i.e.}, mathematical operations, these
pulses are ideal and do not introduce any imperfections. As a result,
this new method is more robust against pulse imperfections and improves
the DD performance significantly. Here, we give a theoretical analysis
of this scheme and show experimentally that it performs better than
the standard CDD method when applied to a single qubit interacting
with a pure dephasing environment - a typical situation for many QIP
implementations \cite{Ladd2010}.

\section{The System}

We consider a single qubit $\hat{S}$ as the system that is coupled
to a bath. The free evolution Hamiltonian is 
\begin{equation}
\widehat{\mathcal{H}}_{f}=\widehat{\mathcal{H}}_{SE}+\widehat{\mathcal{H}}_{E},
\end{equation}
in a suitable rotating frame of reference that is on resonance with
the system qubit \cite{Abragam}. $\widehat{\mathcal{H}}_{E}$ is
the environment Hamiltonian and 
\begin{equation}
\widehat{\mathcal{H}}_{SE}=\sum_{\beta}\left(b_{z}^{\beta}\hat{E}_{z}^{\beta}\hat{S}_{z}+b_{y}^{\beta}\hat{E}_{y}^{\beta}\hat{S}_{y}+b_{x}^{\beta}\hat{E}_{x}^{\beta}\hat{S}_{x}\right)\label{eq:SEintercation}
\end{equation}
 is a general system-environment (SE) interaction. The operators $\hat{E}_{u}^{\beta}$
are environment operators and $b_{u}^{\beta}$ the SE coupling strengths.
The index $\beta$ runs over all modes of the environment. Dephasing
effects come from the interaction that affects the $z$ component
of the spin-system operator, and spin-flips and/or polarization damping
are produced trough the $x$ and/or $y$ operators. We will discuss
our method in a general SE interaction context, but the experimental
results were performed on a spin-system coupled with a spin-bath.
The SE interaction is given by a heteronuclear spin-spin interaction
that effects a pure dephasing. In general, this type of interaction
is naturally encountered in a wide range of solid-state spin systems,
for example in nuclear mangetic resonance (NMR) \cite{Carr1954,Meiboom1958,Alvarez2010,Ajoy2011},
electron spins in diamonds \cite{Ryan2010}, electron spins in quantum
dots \cite{Hanson2007}, donors in silicon \cite{Kane98}, etc. In
other cases, when the system and environment have similar energies,
the SE interaction can include terms along the $x$, $y$ and $z$
axis.

\section{CDD with real and virtual pulses\label{sec:CDD-with-virtual}}

\subsection{CDD}

Concatenated DD (CDD) is a scheme for improving the efficiency of
a DD sequence \cite{Khodjasteh2005,Khodjasteh2007} by recursively
concatenating lower order sequences CDD$_{n-1}$ into a higher-order
sequence CDD$_{n}$ by inserting CDD$_{n-1}$ blocks into the delays
of a generating sequence 
\begin{equation}
\mathrm{CDD}_{n}=C_{n}=C_{n-1}\hat{X}C_{n-1}\hat{Y}C_{n-1}\hat{X}C_{n-1}\hat{Y},\label{cdd}
\end{equation}
where $C_{0}=\tau$ is a free evolution period and $\hat{X}$ and
$\hat{Y}$ are $\pi$-pulses of the generating sequence. $\mathrm{CDD}_{1}=C_{1}=XY4$
consists of four rotations around the $x$- and $y$-axes. Its pulse
sequences is given by $XY4=\tau\mbox{-}\hat{X}\mbox{-}\tau\mbox{-}\hat{Y}\mbox{-}\tau\mbox{-}\hat{X}\mbox{-}\tau\mbox{-}\hat{Y}$.
This sequence can decouple SE interactions that include all three
components of the system spin operator \cite{Viola1999} and it mitigates
the effect of pulse errors compared to the older CPMG sequence consisting
of identical pulses \cite{Maudsley1986}. This can be understood by
considering that pulse imperfections convert an Ising-type SE interaction
into an effective general SE interaction \cite{Khodjasteh2007,Alvarez2010,Souza2011a},
which can be partially eliminated by the $XY4$ sequence. In the QIP
community, the $XY4$ sequence is usually referred to as periodic
dynamical decoupling (PDD). Alternatively, we proposed to use the
time symmetric version of $XY4$ \cite{Maudsley1986} in the CDD protocol
because the resulting CDD(s) sequences are are more efficient at supressing
decoherence and pulse error effects \cite{Souza2011,Souza2011b,Souza2011a}.
These symmetric sequences can be written as \cite{Maudsley1986,Souza2011,Souza2011b}
$XY4(\mbox{s})=\tau/2\mbox{-}\hat{X}\mbox{-}\tau\mbox{-}\hat{Y}\mbox{-}\tau\mbox{-}\hat{X}\mbox{-}\tau\mbox{-}\hat{Y}\mbox{-}\tau/2$
and 
\begin{multline}
\mbox{CDD}(\mbox{s})_{n}=C(s)_{n}=\\
=\sqrt{C(s)_{n-1}}\hat{X}C(s)_{n-1}\hat{Y}C(s)_{n-1}\hat{X}C(s)_{n-1}\hat{Y}\sqrt{C(s)_{n-1}}.
\end{multline}

\begin{figure*}
\includegraphics[bb=0bp 45bp 842bp 595bp,clip,width=0.9\textwidth]{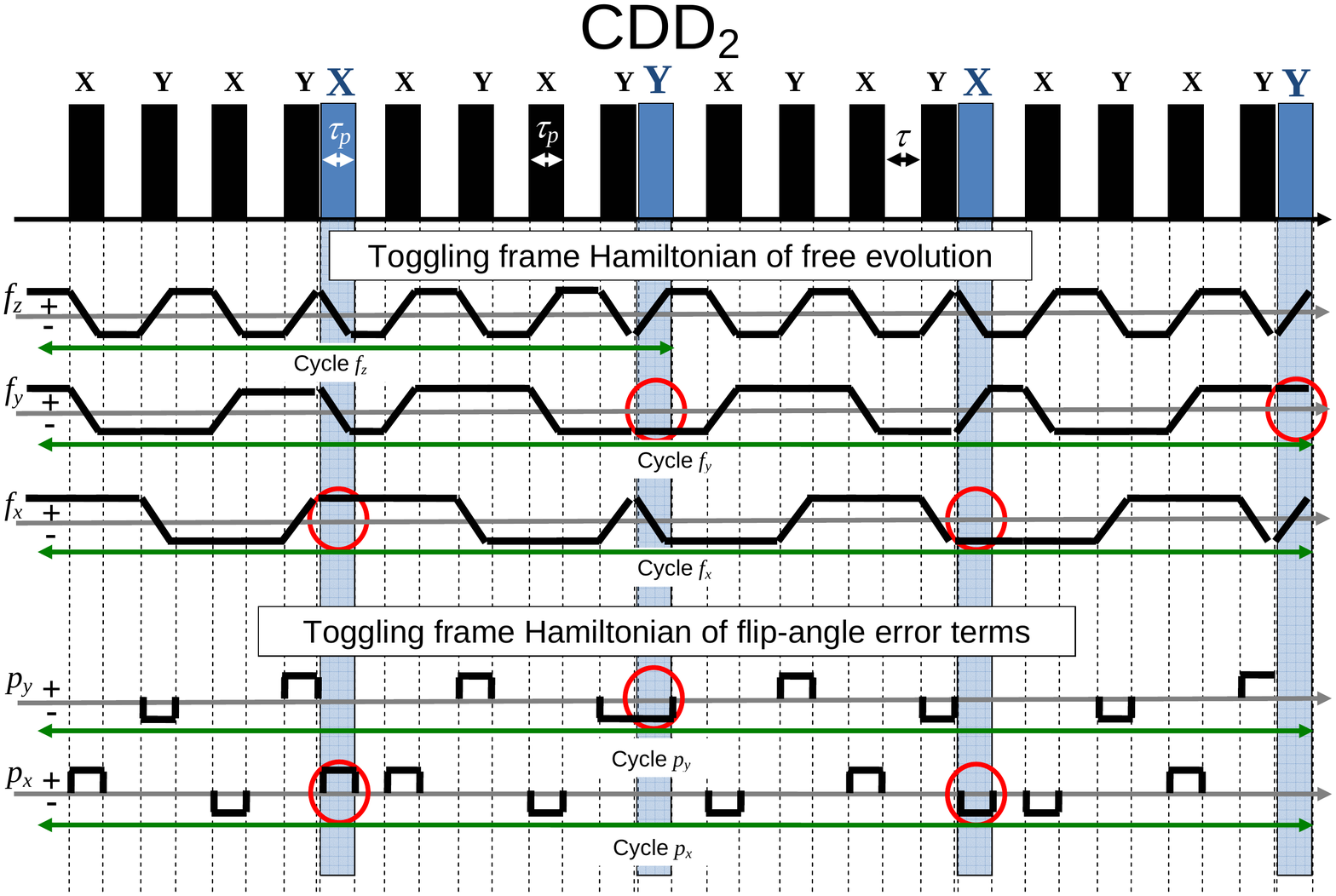}

\includegraphics[bb=0bp 53bp 842bp 595bp,clip,width=0.9\textwidth]{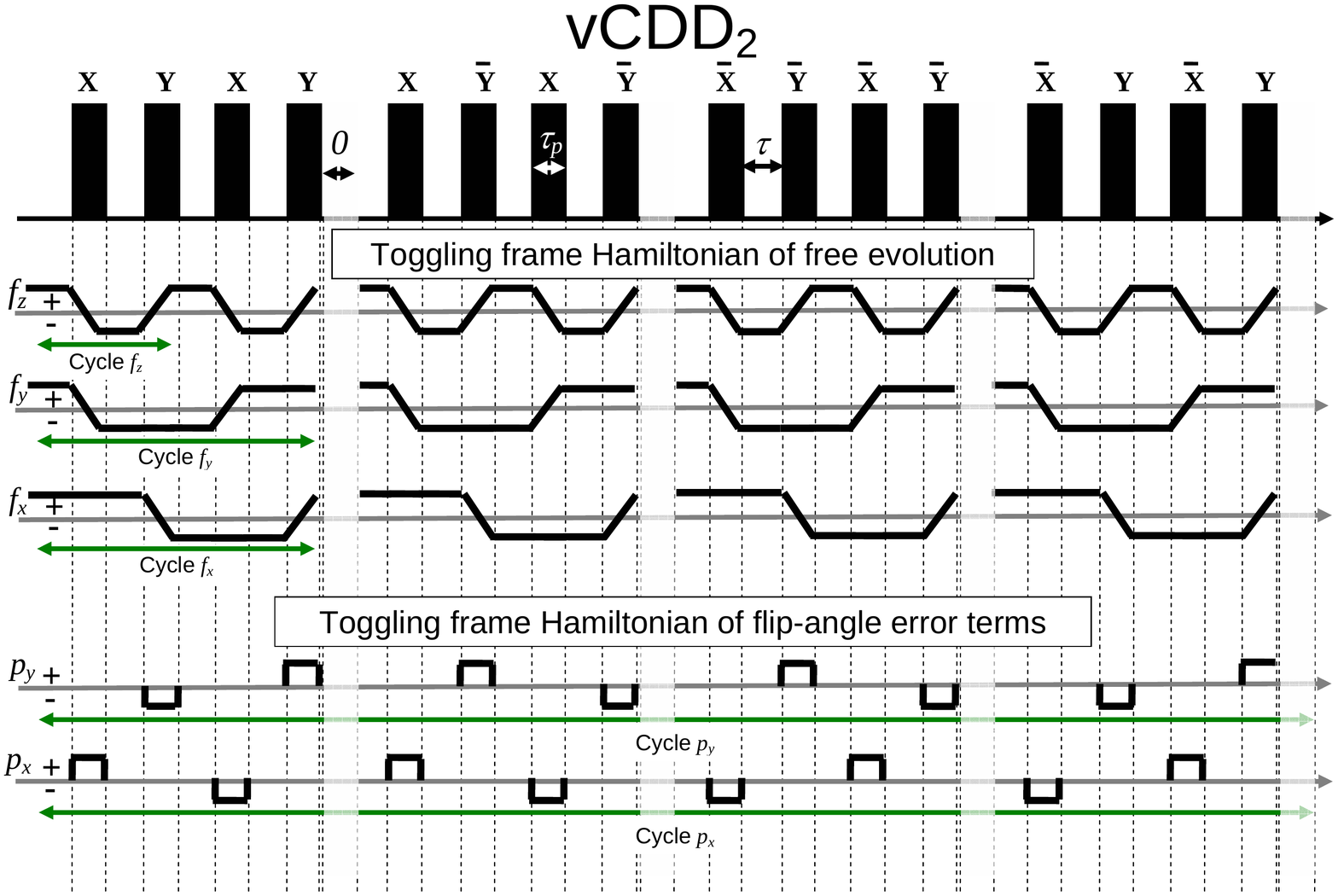}

\caption{(Color online) CDD$_{2}$ and vCDD$_{2}$ pulse sequence schemes.
The black solid boxes represent the DD $\pi$-pulses of the inner
sequences with their respective phases. The gray (blue) boxes are
the $\pi$-pulses of the generating sequence for CDD, while for vCDD
they are virtual and appear as a transparent white stripe of zero
duration. The toggling frame Hamiltonians are represented by the respective
signs of the different terms proportional to the $\hat{S}_{x}$, $\hat{S}_{y}$
and $\hat{S}_{z}$ components. Sign changes during the pulses are
representedby diagonal lines \textbackslash{} or /. In the CDD$_{2}$
scheme, the terms marked by circles are compensated only at the end
of the complete cycle, but the vCDD scheme compensates all terms over
the basic 4-pulse cycle. The toggling frame Hamiltonians of the free
evolution interaction for vCDD$_{2}$ are the same for all blocks
of the inner sequence, i.e., equal to those of the $XY4$ sequence.
The toggling frame Hamiltonian of the flip-angle error terms of vCDD
is equal to that of the CDD scheme with ideal pulses.}
\label{Flo:CDDvCDDscheme}
\end{figure*}

The square root $\sqrt{C(s)_{n}}$ represents half of the cycle. Each
level of concatenation reduces the norm of the first non-vanishing
term of the Magnus expansion of the previous level, provided that
the norm was small enough to begin with \cite{Khodjasteh2005,Khodjasteh2007}.
This reduction comes at the expense of an increase of the cycle time
by a factor of four. The average Hamiltonian can be calculated in
the toggling frame. If the pulses generate ideal $\pi$-rotations,
this can be seen as a sign change of different terms of the SE interaction
(\ref{eq:SEintercation}). The top panel of Fig. \ref{Flo:CDDvCDDscheme}
shows the CDD$_{2}$ scheme and it shows the sign changes of the different
terms of the SE interaction in the toggling frame. The parameters
$f_{u}$ with $u=x,y,z$ represent the signs of the terms of Eq. (\ref{eq:SEintercation})
that are proportional to $\hat{S}_{x}$, $\hat{S}_{y}$ and $\hat{S}_{z}$,
respectively, in the toggling frame.

\subsection{Effect of pulse imperfections in CDD}

Since the precision of any real pulse is finite, they generate an
evolution that differs from the ideal one. If many pulses are applied
in sequence, these errors can accumulate and seriously reduce the
fidelity of the evolution \cite{Alvarez2010,Souza2011,Souza2011a,Wang2011,Xiao2011,Wang2010},
unless the sequence of operations is designed in such a way that the
errors from different pulses compensate each other \cite{Souza2011,Souza2011a}.
One kind of error of non-ideal control pulses is their finite duration,
which implies a minimum achievable cycle time. The effects introduced
by finite pulse lengths have been considered in different theoretical
works \cite{Viola2003,Khodjasteh2007,Hodgson2010}. These works predict
that high order CDD sequences can lose their advantages when the delays
between pulses or pulse length are strongly constrained. This is because
the fundamental frequency $2\pi/\tau_{c}$, where $\tau_{c}$ is the
period of the toggling frame function $f(t)$, is lower for the longer
cycle \cite{Ajoy2011}. The efficiency of all DD sequences is reduced
if the noise contains frequency components at the resonance frequencies
of their filter function \cite{Alvarez2011}. This was demonstrated
for UDD sequences, but the analysis is similar for CDD sequences because
the period of the toggling frame sign function $f$ increases with
the concatenation order \cite{Ajoy2011}.

As shown in Fig. \ref{Flo:CDDvCDDscheme}, the toggling frame Hamiltonian
for one of the components is not affected by the pulses of the generating
sequence (marked by circles). Due to the finite duration of the pulses,
this represents an additional contribution to the average Hamiltonian,
which is only compensated by the second pulse of the generating sequence
with the same rotation axis half a period later. Full compensation
of these additional terms is achieved at the end of the complete (higher-order)
cycle.

Generally more important than their finite duration are imperfections
of the pulses. In most cases, the dominant cause of errors is a deviation
between the ideal and the actual amplitude of the control fields.
This results in a rotation angle that deviates from $\pi$, typically
by a few percent. The propagator for the $\pi$ pulses including this
error is $e^{-i\left(\pi+\Delta\omega_{p}\tau_{p}\right)\hat{S}_{\phi}}$,
where $\Delta\omega_{p}$ is the error on the control field amplitude.
In the toggling frame Hamiltonian, the ideal part of this propgator,
$e^{-i\pi\hat{S}_{\phi}}$, vanishes, but the error term $e^{-i\Delta\omega_{p}\tau_{p}\hat{S}_{\phi}}$
remains and contributes to the average Hamiltonian. The signs of these
terms in the toggling frame are represented as $p_{u}$ in Fig. \ref{Flo:CDDvCDDscheme}.

Another important error occurs when the control field is not applied
on resonance with the transition frequency of the qubit. This off-resonance
error adds a term $f_{z}\Delta_{z}\hat{S}_{z}$ to the toggling-frame
Hamiltonian.

\begin{figure*}
\includegraphics[bb=10bp 215bp 440bp 600bp,clip,width=0.97\columnwidth]{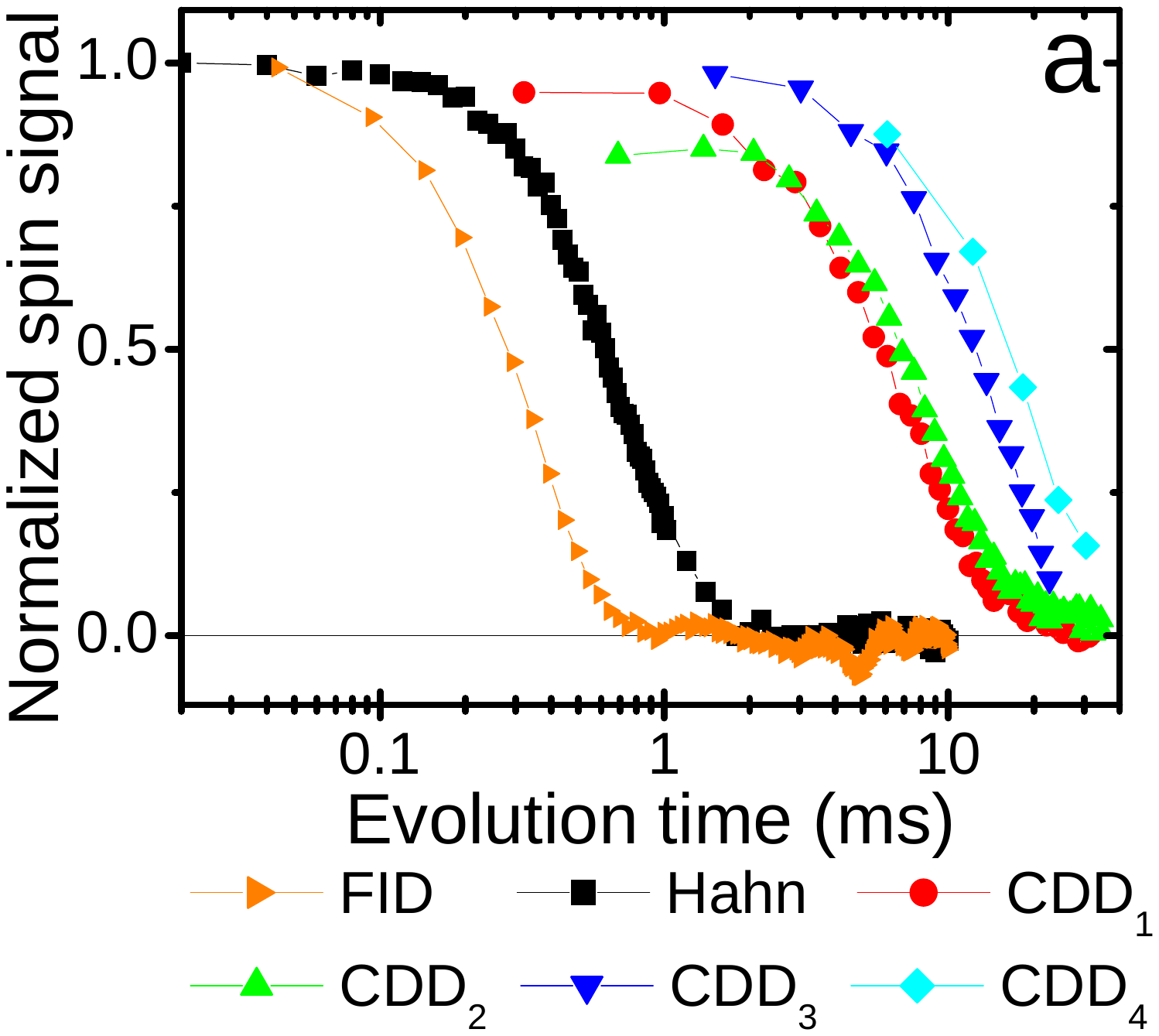}\includegraphics[bb=350bp 230bp 792bp 612bp,clip,width=1.03\columnwidth]{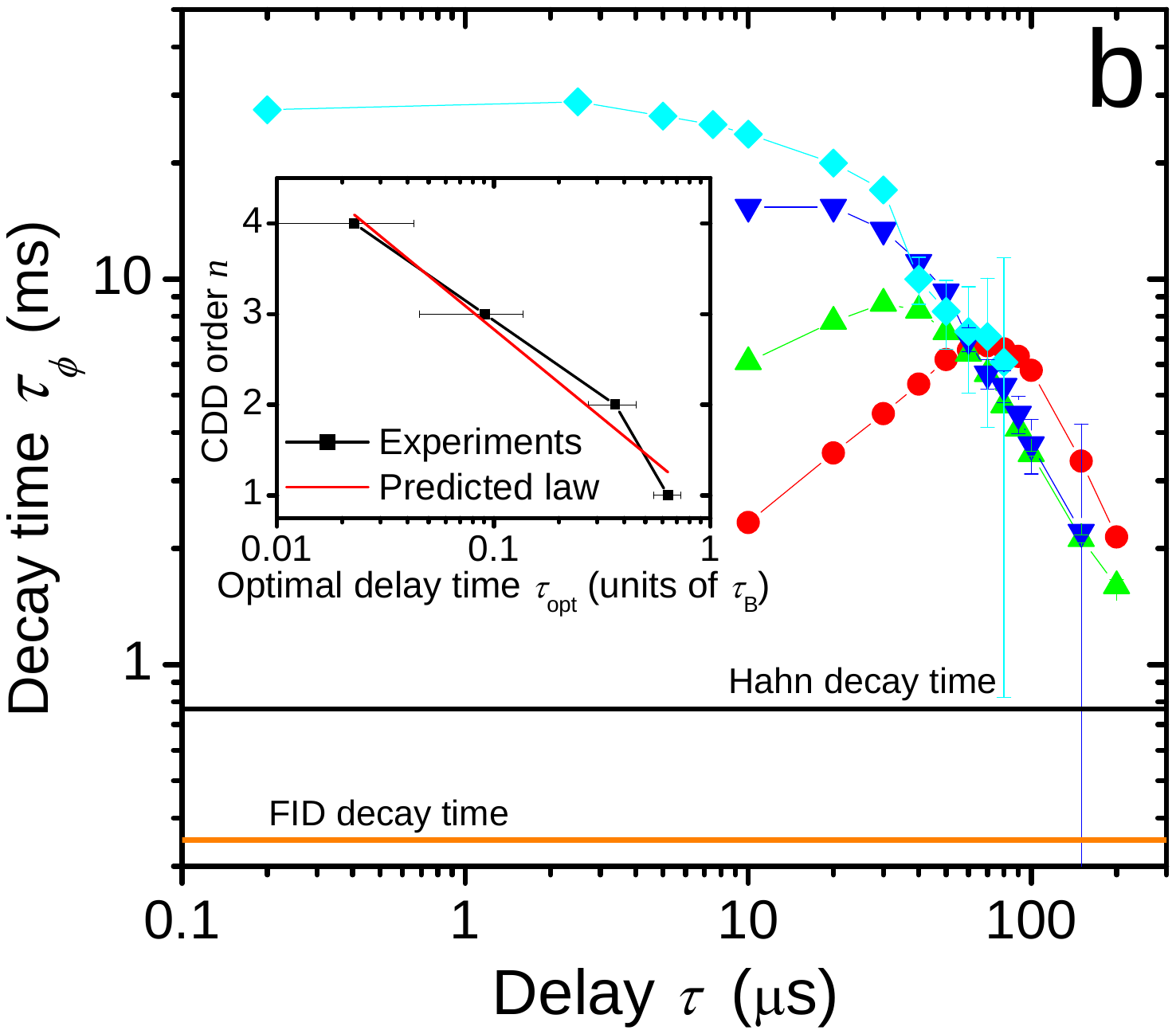}

\caption{(Color online) Decays of coherence under the influence of different
DD sequences. (a) Normalized spin-signal decay of the echo trains
of different CDD sequences, the Hahn echo decay \cite{Hahn1950} and
the free evolution (FID). The CDD decay curves are plotted for their
optimal delays between pulses that are given when the curves of panel
b have a maximum. (b) Decay times of different CDD sequences for different
delays between pulses. The optimal delay time is defined when the
decay is a maximum. The inset shows its dependence as a function of
the CDD order matching with theoretical predictions \cite{Khodjasteh2007}.}
\label{Flo:CDDFigure}
\end{figure*}

The $XY4$ sequence cancels these errors in zeroth order), independent
of the initial condition \cite{Maudsley1986,Gullion1990}. As a result,
the performance of this sequence is quite symmetric with respect to
the initial state in the $xy$-plane and the average decay times are
significantly longer than with non-robust sequences \cite{Alvarez2010,Ryan2010,Souza2011,Souza2011a}.
The concatenation scheme proposed by Khodjasteh and Lidar \cite{Khodjasteh2005,Khodjasteh2007}
improves the decoupling performance and the tolerance to pulse imperfections
\cite{Alvarez2010,Souza2011}. However, the finite duration of the
pulses and constrained delays between pulses result in the existence
of optimal levels of concatenation \cite{Alvarez2010,Souza2011},
with decreasing performance for higher level sequences. This can be
seen in Fig. \ref{Flo:CDDFigure}, where decay curves are ploted for
different DD sequences, including the free evolution decay, the Hahn
echo decay \cite{Hahn1950} and different orders of CDD for their
optimal delay between pulses. Panel b shows the decay times for different
CDD sequences and delays $\tau$ between pulses. For each sequence,
the decoherence time reaches a maximum; for delays shorter than the
optimal value, the pulse errors dominate. The relation between the
optimal delay time and the its CDD order is plotted in the inset of
Fig. \ref{Flo:CDDFigure}b). The experimental dependence agrees remakrably
well with the predicted curve \cite{Khodjasteh2007}.

\subsection{CDD with virtual pulses}

In Ref. \cite{Alvarez2010}, we suggested to improve the concatenation
scheme by compensating the pulse errors of the generating sequence
(gray (blue) boxes, Fig. \ref{Flo:CDDvCDDscheme}) before the end
of the complete cycle. Looking into the details of the toggling frame
Hamiltonians, we can see that at each concatenation level, the $XY4$
generating sequence (gray (blue) boxes on the top panel of Fig. \ref{Flo:CDDvCDDscheme})
additional pulse errors are introduced that are only compensated at
the end of the complete cycle. As a result, the properties of the
real CDD sequence deviate strongly from that of the ideal sequence.

Here, we show how these additional pulse errors can be completely
avoided by using virtual (and thus ideal) rotations for the generating
sequence instead of the real ones. To motivate the idea, we consider
the first pulse of the generating sequence and the subsequent pulses
of the cycle from the lower order sequence. The corresponding evolution
operator can be written 

\begin{eqnarray}
\ldots\left(\hat{Y}\hat{X}\hat{Y}\hat{X}\right)\hat{X}\ldots & = & \ldots\hat{X}\left(\hat{\overline{Y}}\hat{X}\hat{\overline{Y}}\hat{X}\right)\ldots,
\end{eqnarray}
where the pulse sequence is read from right to left. The bar over
the $X$ and $Y$ means that the sense of rotation is reversed for
those pulses. The second form corresponds to a modified $XY4$ cycle,
followed by the $\pi_{x}$ pulse of the generating sequence. In the
modified cycle, the direction of rotation of the $y$-pulses has been
inverted. We distinguish this modified cycle from the original cycle
by writing them as $vC_{n-1}(X,\bar{Y})$ and $vC_{n-1}(X,Y)$, respectively.
Similarly, the subsequent cycles become $vC_{n-1}(\bar{X},\bar{Y})$
and $vC_{n-1}(\bar{X},Y)$. As the pulses of the generating sequence
are thus moved to the end of the cycle, they cancel and can be omitted
completely. In this sense, we have replaced these pulses by ``virtual
pulses'' corresponding to phase changes of the pulses in the inner
sequence. The resulting sequence, which is shown in Fig. \ref{Flo:CDDvCDDscheme},
can be written recursively as 
\begin{align}
\mbox{vCDD}_{1}(X,Y) & =XY4\label{vcdd1}
\end{align}
\begin{multline}
\mathrm{vCDD}_{n}(X,Y)=vC_{n}(X,Y)=\\
vC_{n-1}(X,Y)\mbox{-}vC_{n-1}(X,\bar{Y})\mbox{-}vC_{n-1}(\bar{X},\bar{Y})\mbox{-}vC_{n-1}(\bar{X},Y).\label{eq:vcdd}
\end{multline}
Similarly we can obtain its time-symmetric version. 

As shown in Fig. \ref{Flo:CDDvCDDscheme}, the toggling frame Hamiltonian
generated by this sequence differs from that of the original CDD sequence.
As shown in the lower part of the figure, the function $f$ has for
each 4-pulse block the same time dependence as for the $XY4$ sequence.
The terms marked by circles in the upper part of the figure are missing
in the lower part; accordingly, the average of the $f_{u}$ vanishes
over each block of the inner sequence. Similarly, the pulse error
contributions $p_{u}$ do not have contributions from the generating
sequence and therefore also compensate over each lower order cycle.
In lowest order average Hamiltonian, the vCDD sequences therefore
compensate all errors over a single $XY4$ cycle, while the corresponding
time for CDD$_{N}$ is $4^{N}$ times longer. For the vCDD sequence,
the lowest frequency of the filter function is therefore always $2\pi/\tau_{1}$,
where $\tau_{1}$ is the duration of the CDD$_{1}=XY4$ cycle. In
contrast to that, the fundamental frequency of the CDD$_{n}$ sequence
decreases with $1/4^{n}$, which can make it sensitive to low-frequency
noise with high amplitudes, such as frequency offsets and errors of
control fields.

The change in the toggling frame Hamiltonian effected by the pulses
of the generating sequence of CDD can of course also be a desired
property, since it compensates higher-order terms of the average Hamiltonian,
including cross-terms between pulse imperfections and environmental
contributions. Some of these effects are also present in the vCDD
scheme, since the non-vanishing higher-order average Hamiltonians
of the different blocks are not identical. The concatenation scheme
is designed to compensate them over the full cycle. A detailed discussion
of these higher-order contributions is beyond the scope of this paper
and probably not feasible without considering specific system parameters.
Instead, we compare the two schemes experimentally.

\section{Experimental performance comparison}

\subsection{System and setup}

We experimentally implemented the new vCDD scheme and compared its
performance to that of the normal CDD scheme. The experiments were
performed on a polycrystalline adamantane sample using a home-built
solid state NMR spectrometer with a $^{\text{1}}$H resonance frequency
of 300 MHz. Our system qubits are the $^{13}$C nuclear spins of the
adamantane molecule, which contains two nonequivalent carbon atoms.
Under our conditions, they have similar dynamics. Here, we present
the results from the CH$_{2}$ carbon. Working with a natural abundance
sample (1.1 \% $^{13}$C), the interaction between the $^{13}$C-nuclear
spins can be neglected. The main mechanism for decoherence is the
interaction with the neighboring proton spins. As discussed before,
this interaction generates pure dephasing. This interaction is not
static, since the dipole-dipole couplings within the proton bath cause
flip-flops of the protons coupled to the carbon. The $\text{\ensuremath{\pi}}$
pulses for DD were applied on resonance with the $^{13}$C spins.
Their radio-frequency (RF) field of $\approx2\pi\times50$ kHz gives
a $\text{\ensuremath{\pi}}$-pulse length $\text{\ensuremath{\tau_{p}}}$
between $10\mu\mbox{s}$ and $10.6\mu\mbox{s}$. The measured RF field
inhomogeneity was about 10\%.

\subsection{vCDD and CDD under optimal conditions}

\begin{figure}
\includegraphics[bb=50bp 30bp 700bp 550bp,clip,width=0.95\columnwidth]{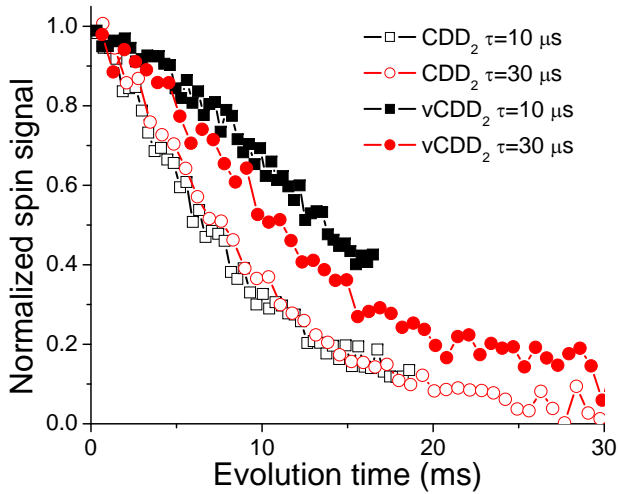}

\caption{(Color online) Normalized spin signal decays for vCDD$_{2}$ and CDD$_{2}$
for different delays $\tau$ .}
\label{Flo:signaldecayVcddvsCDD}
\end{figure}

Figure \ref{Flo:signaldecayVcddvsCDD} compares the decay of the spin
signal for the asymmetric versions of CDD$_{2}$ and vCDD$_{2}$ for
two different pulse spacings $\tau$. For the vCDD$_{2}$-sequence,
the decay is clearly slower; the $1/e$ decay times are 17 ms and
14.6 ms for the two delays, compared to 8.9 ms and 10.1 ms for the
CDD2 sequence. 

\begin{figure}
\includegraphics[bb=40bp 20bp 700bp 550bp,clip,width=0.9\columnwidth]{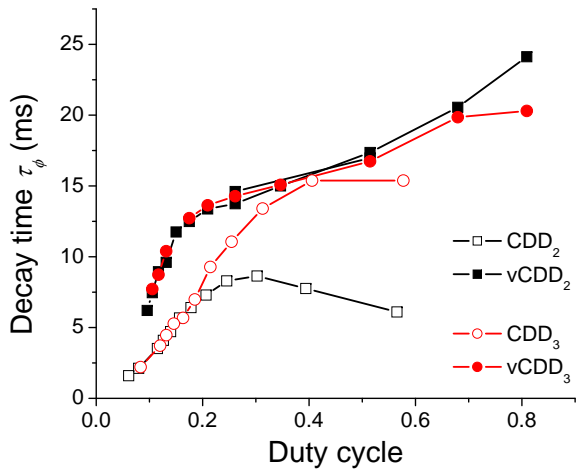}

\caption{(Color online) Decay times for vCDD and CDD of order 2 and 3 as a
function of the duty cycle.}
\label{Flo:CDDvsvCDDvsduty}
\end{figure}

Figure \ref{Flo:CDDvsvCDDvsduty} shows the decay times for different
asymmetric CDD and vCDD orders obtained for different duty cycles,
\emph{i.e.}, the ratio between the irradiation time $N_{p}\tau_{p}$
over the total time $(N_{p}\tau_{p}+N\tau)$, where $N_{p}$ is the
number of pulses in a cycle and $N$ the number of delays. $\tau_{p}$
was kept fixed and we varied the delay $\tau$ between the pulses.
Comparing the curves for the two schemes, we find that vCDD performs
better than the CDD sequences for all duty cycles (delays). While
the CDD performance changes as a function of the order, the difference
between the two vCDD sequences is not significant. The difference
beween the symmetric and asymmetric version of vCDD also was not significant.
This suggests that the 2$^{nd}$ order achieves already the optimal
DD performance for our experimental conditions. The observed performance
is also very similar to that of the KDD sequence measured in an earlier
study (see Ref. \cite{Souza2011} for details). Both, the vCDD and
the CDD sequences perform symmetrically for initial conditions in
the $xy$-plane.


\subsection{Effect of pulse errors}

\begin{figure}
\includegraphics[bb=0bp 0bp 580bp 792bp,clip,width=1\columnwidth]{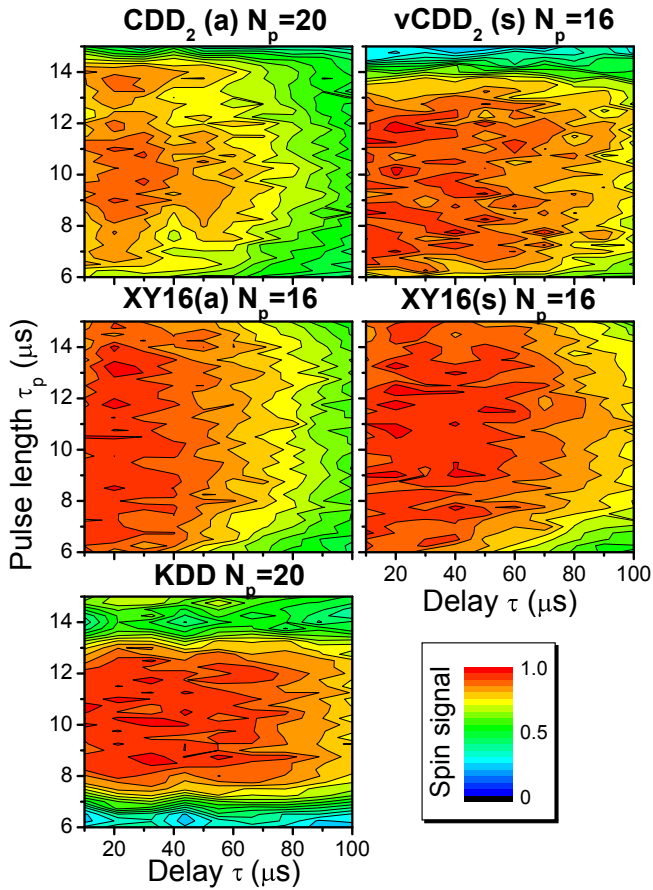}

\caption{(Color online) Normalized spin-signal after 1 cycle of different DD
sequences as a function of the pulse-length of the DD pulses and the
delay between them. The labels (a) and (s) refers the the asymmetric
and symmetric version of the sequences.}
\label{Flo:DDseqvsflipangleerror1}
\end{figure}

\begin{figure}
\includegraphics[bb=0bp 0bp 580bp 792bp,clip,width=1\columnwidth]{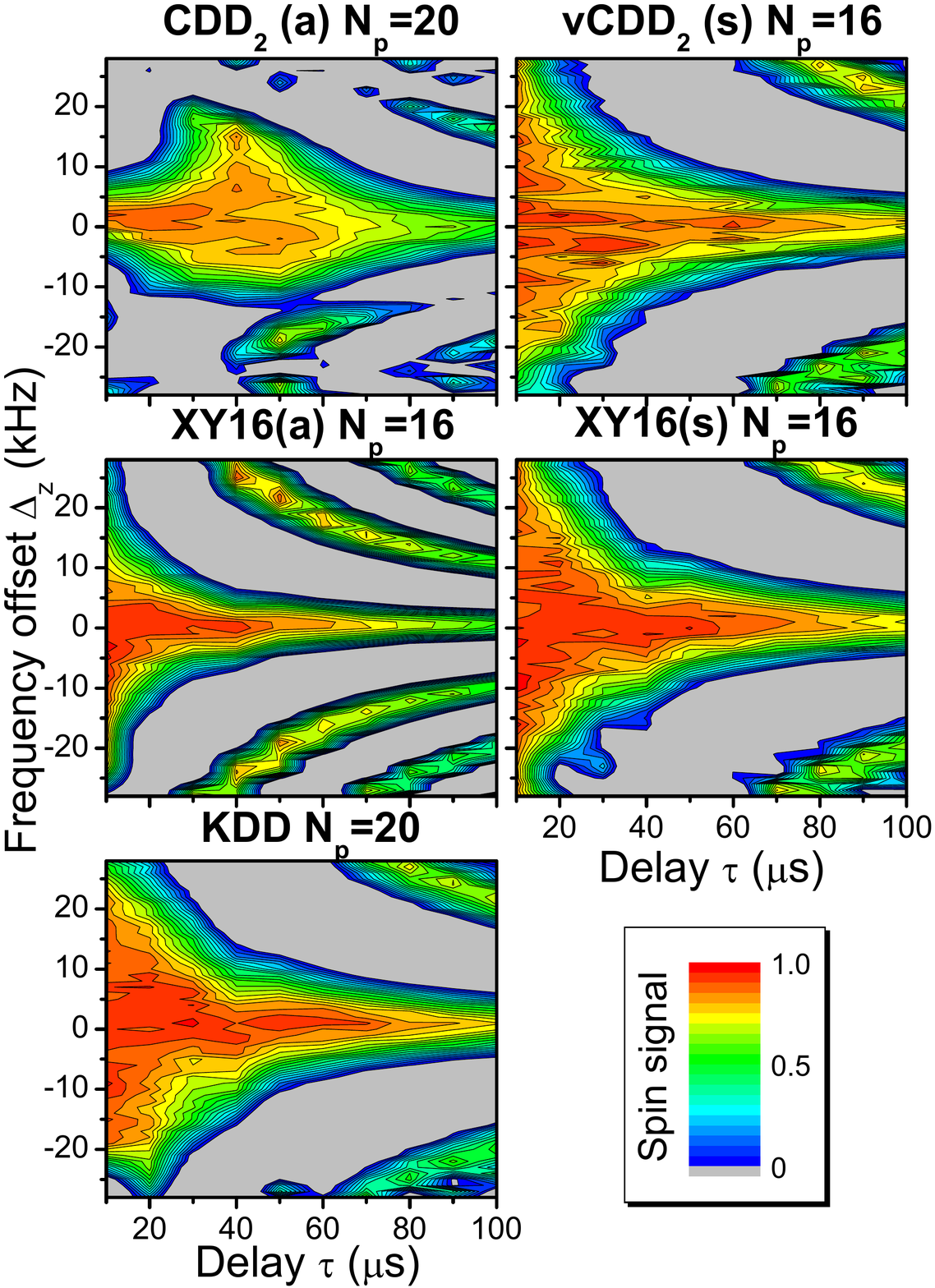}

\caption{(Color online) Normalized spin-signal after 1 cycle of different DD
sequences as a function of the RF pulse frequency of the DD pulses
and the delay between them. The labels (a) and (s) refers the the
asymmetric and symmetric version of the sequences.}
\label{Flo:DDseqvsoffset1}
\end{figure}

Under normal experimental conditions, we cannot see any difference
between the vCDD scheme and other robust sequences like KDD and $XY16$
\cite{Alvarez2010,Souza2011,Souza2011a,Souza2011b}. To quantitate
this robustness we also tested the performance of the sequence against
artificially added pulse errors. We compared CDD with vCDD and with
the optimal sequences obtained in previous works, \emph{i.e.}, $XY16$
and KDD \cite{Souza2011,Souza2011a,Souza2011b}. Figures \ref{Flo:DDseqvsflipangleerror1}
and \ref{Flo:DDseqvsoffset1} compare the spin signal after one cycle
of the respective sequence for different pulse errors. Fig. \ref{Flo:DDseqvsflipangleerror1}
shows the surviving spin polarization as a function of the pulse duration
(and thus of the flip angle) and the delay between the pulses. The
number of pulses per cycle is not exactly the same for the different
sequences (16 for vCDD$_{2}$ and $XY16$ vs. 20 for CDD$_{2}$ and
KDD), but we consider this to be sufficiently similar to allow a rough
comparison. For all sequences, there is little correlation between
the flip angle error and the delay between the pulses. This is a consequence
of the fact that the terms in the propagator that involve the flip-angle
error are proportional to the pulse width $\tau_{p}$ but independent
of the pulse separation $\tau$. 

Figure \ref{Flo:DDseqvsoffset1} shows similar data, but here we introduced
an artificial offset error $\Delta_{z}$ rather than a flip-angle
error. In this case, there is a strong correlation between the effect
of the offset and the delay between the pulses. This is expected,
because an offset error generates an extra dephasing term in the propagator
that generates an additional precession by an angle $\Delta_{z}\tau$.
Without the SE interaction or another source of errors (like flip-angle
error inhomogenity), we do not expect a significant dependence on
$\tau$, because the offset is static and can be completly refocused
with DD. Our real system has a bath correlation time of $\approx100$
$\mu$s \cite{Alvarez2010,Ajoy2011}, which explains the observed
decay for cycle times of this order.
\begin{figure}
\includegraphics[bb=0bp 0bp 580bp 300bp,clip,width=1\columnwidth]{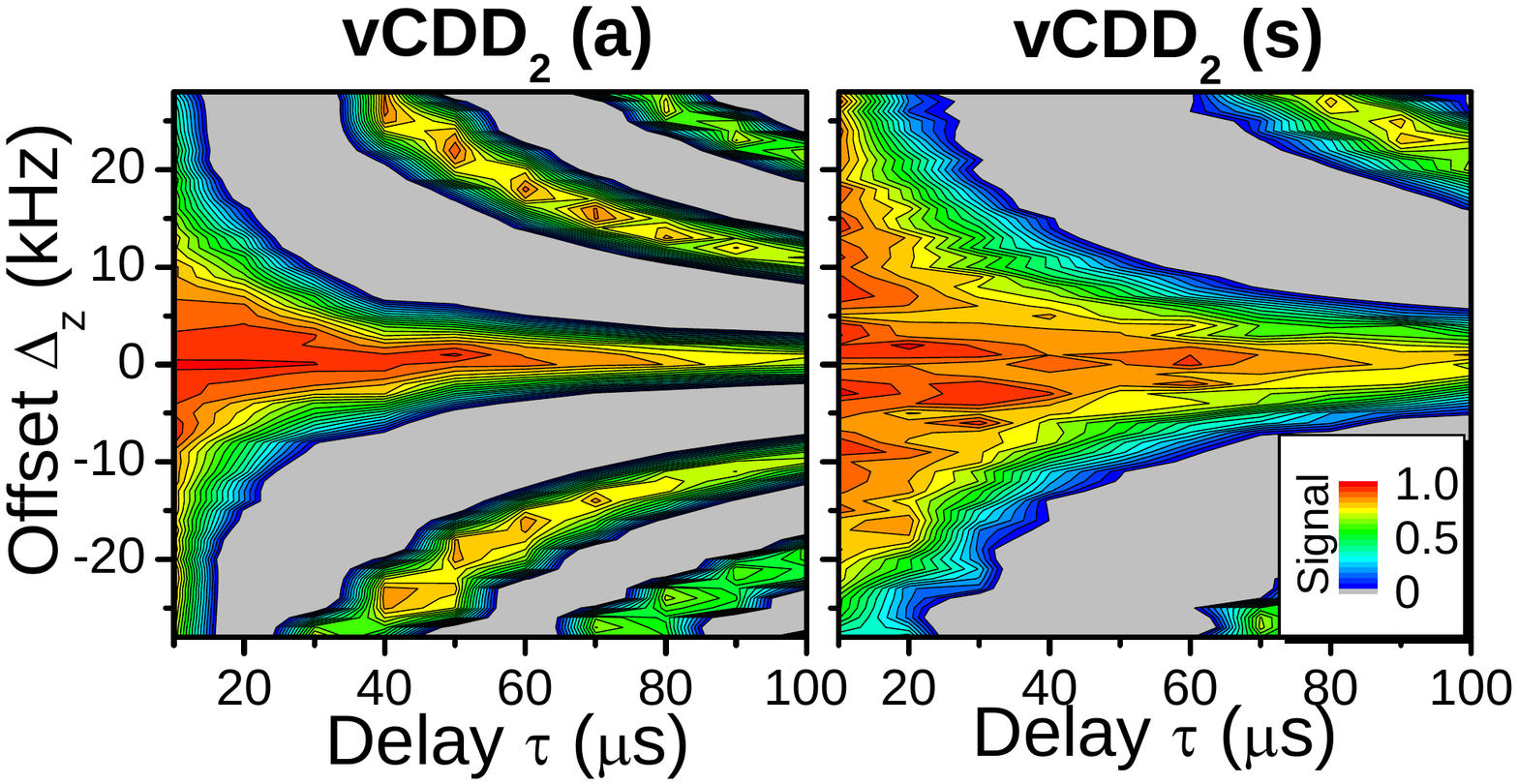}

\caption{(Color online) Normalized spin-signal after 1 cycle for the symmetric
(s) and asymmetric (a) version of vCDD$_{2}$ as a function of the
offset frequency of RF pulse of the DD pulses and the delay between
them. Both sequences have the same number of pulses and cycle time.}

\label{Flo:vCDD(a)_vs_(s)}
\end{figure}

Comparing the standard CDD$_{2}$ with the symmetric version of vCDD$_{2}$
in Fig. \ref{Flo:DDseqvsflipangleerror1}a,b, we can see that the
overall performance of vCDD is better than that of CDD, as expected
by the analysis of section \ref{sec:CDD-with-virtual}. This is because
vCDD is more effective in compensating the flip-angle errors. vCDD
also outperforms CDD in the presence of offset errors (see Fig. \ref{Flo:DDseqvsoffset1}a,b).
Comparing the asymmetric and symmetric version of vCDD as a function
of flip angle error, we observe no significant differences. However,
vCDD(s) clearly outperfroms vCDD(a) in the presence of offset errors
(Fig. \ref{Flo:vCDD(a)_vs_(s)}). Comparing against the other sequences,
vCDD$_{2}$ seems to perform better than KDD as a function of flip-angle
errors. The good performance of $XY16$ is expected because its evolution
operators (symmetric and asymmetric) are equal to the identity operator
as long as spin-bath effects are absent: the sequence is designed
to generate a propagator $U\, U^{\dagger}$, independent of flip-angle
errors. For small delays between pulses, $XY16$ is the most robust
sequence as a function of flip-angle error. Its symmetric version
performs slightly better than the asymmetric version. As a function
of offset error, vCDD$_{2}$(s), KDD and $XY16$(s) behave similarly
and they are more robust than vCDD$_{2}$(a) and $XY16$(a). Note
that the behaviour of the asymmetric version of vCDD$_{2}$ and $XY16$
as a function of offset errors are also similar.

\begin{figure}
\includegraphics[bb=0bp 0bp 580bp 792bp,clip,width=1\columnwidth]{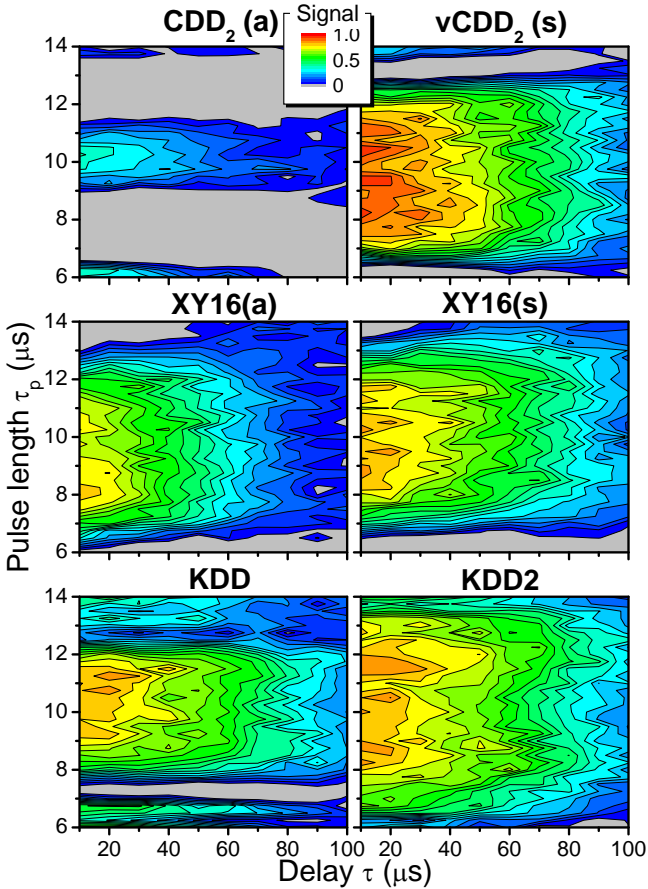}

\caption{(Color online) Normalized spin-signal after about 100 pulses for different
DD sequences as a function of the pulse-length of the DD pulses and
the delay between them. All sequences have 100 pulses except vCDD$_{2}$,
which contains 96. The labels (a) and (s) refers the the asymmetric
and symmetric version of the sequences.}
\label{Flo:DDseqvsflipangleerror100}
\end{figure}

\begin{figure}
\includegraphics[bb=0bp 0bp 580bp 792bp,clip,width=1\columnwidth]{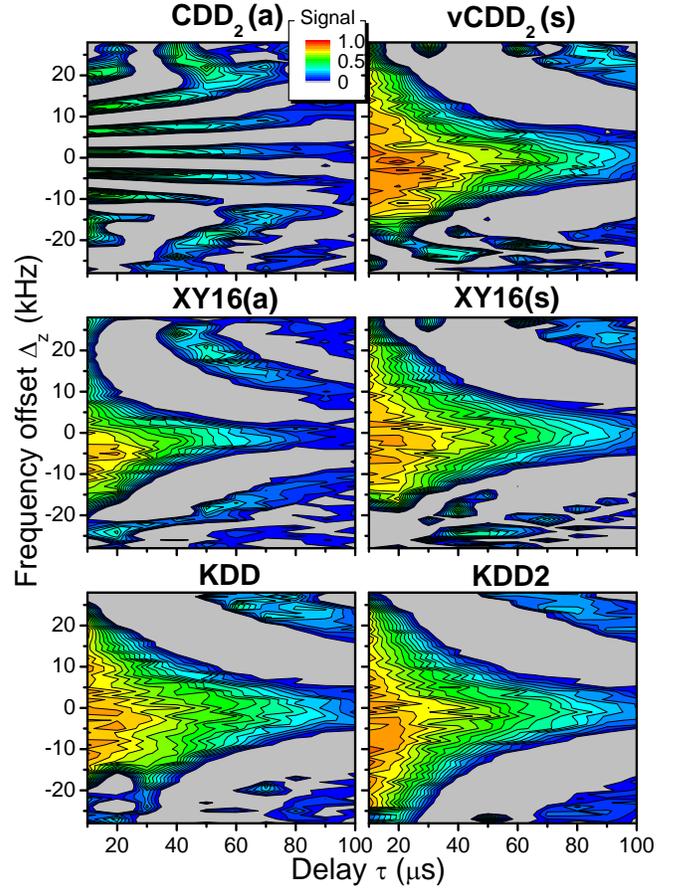}

\caption{(Color online) Normalized spin-signal after about 100 pulses for different
DD sequences as a function of the RF frequency of the DD pulses and
the delay between them. All sequences have 100 pulses except vCDD$_{2}$,
which contains 96. The labels (a) and (s) refers the the asymmetric
and symmetric version of the sequences.}
\label{Flo:DDseqvsoffseterror100}
\end{figure}

To amplify the effect of pulse imperfections, we also performed experiments
with $\approx$100 pulses as a function of the delay between the pulses
and added specific pulse imperfections (Figs. \ref{Flo:DDseqvsflipangleerror100}
and \ref{Flo:DDseqvsoffseterror100}). Under these conditions, also
the accumulated exposure to the spin-bath is longer. Cleary now the
vCDD sequence outperforms always the CDD version for every condition.
As a function of flip angle error, the performance of vCDD(s) and
vCDD(a) is comparable (vCDD(a) is not shown). The vCDD perfomance
as a function of flip angle error is even better than KDD and comparable
to the $XY16$(s) . As a function of offset error, the performance
of vCDD$_{2}$(s) is comparable to KDD and $XY16$(s); however in
this case vCDD$_{2}$(a) is less robust (not shown in the figure). 


\section{Other generating sequences}

The concept introduced here can can not only be applied to the $XY4$
sequence but also to other generating sequences, such as the KDD sequence
\cite{Souza2011}. KDD was inspired from a sequence of adjacent $\pi$
pulses that combine to a robust $\pi$ pulse \cite{Tycko1985}
\begin{equation}
\Pi_{\phi}=\pi_{\phi+30}-\pi_{\phi+0}-\pi_{\phi+90}-\pi_{\phi+0}-\pi_{\phi+30}.\label{eq:knill_pulse}
\end{equation}
The decoupling sequence is obtained first by introducing delays between
the individual pulses: \cite{Souza2011}
\begin{equation}
\Pi_{\phi}(\tau)=\pi_{\phi+30}\mbox{-}f_{\tau}\mbox{-}\pi_{\phi+0}\mbox{-}f_{\tau}\mbox{-}\pi_{\phi+90}\mbox{-}f_{\tau}\mbox{-}\pi_{\phi+0}\mbox{-}f_{\tau}\mbox{-}\pi_{\phi+30}.\label{eq:knill_buildingblock sequebce}
\end{equation}
The lower indexes denote the pulse phase, i.e. the orientation of
the rotation axis in the $xy$-plane. If we use $XY4$ as the (virtual)
generating sequence and $\Pi_{\phi}(\tau)$ as building blocks, we
arrive at
\begin{equation}
\mbox{KDD}=\Pi_{X}(\tau)-\Pi_{Y}(\tau)-\Pi_{X}(\tau)-\Pi_{Y}(\tau),
\end{equation}
which we introduced and tested in \cite{Souza2011}.

If we use the sequence (\ref{eq:knill_pulse}) instead of $XY4$ as
the (virtual) generating sequence, we obtain a new sequence 
\[
\mbox{KDD}2=\left[\Pi_{30}(\tau)-\Pi_{0}(\tau)-\Pi_{90}(\tau)-\Pi_{0}(\tau)-\Pi_{30}(\tau)\right]^{2}.
\]
As indicated by the square after the bracket, the complete cycle consists
of 50 pulses. Iterations to higher order are of course possible but
will not be covered here.

In Fig. \ref{Flo:DDseqvsflipangleerror100} and \ref{Flo:DDseqvsoffseterror100},
we also show the experimental performance of this new sequence, together
with the sequences discussed earlier. We clearly see that this new
sequene is extremely robust and outperfoms all other sequences.

\section{Conclusions}

We have presented a novel method for concatenated dynamical decoupling:
for the generating sequence, we use virtual rotations instead of physical
control operations. Since these rotations are ideal, our new scheme
avoids introducing additional pulse imperfections, reduces the power
deposition on the system and makes the resulting sequences more robust.
As a result of the reduced number of control operations, the toggling
frame Hamiltonian has a different time dependence than in the standard
CDD scheme. We have tested two different expansion schemes based on
these virtual rotations, called vCDD and KDD2. Both types of sequences
have proved to be very robust under our experimental conditions. It
will be interestig to see if these results can be reproduced in other
systems.
\begin{acknowledgments}
Acknowledgments.---We aknowledge discussion with Daniel Lidar and
Gregory Quiroz. This work is supported by the DFG through Su 192/24-1. 
\end{acknowledgments}
\bibliographystyle{apsrev} \bibliographystyle{apsrev}
\bibliography{DD}

\end{document}